\documentclass[aps,prl,twocolumn,nofootinbib,longbibliography]{revtex4-1}

\usepackage{amsmath, empheq,braket,amsthm,amssymb,graphics,amsfonts,array,color,subfigure, mathrsfs}

\definecolor{myurlcolor}{rgb}{0,0,0.7}
\definecolor{myurlcolor1}{rgb}{0,0.7,0.1}
\definecolor{myrefcolor}{rgb}{0,0,0.7}
\usepackage[unicode=true,pdfusetitle, bookmarks=true,bookmarksnumbered=false,bookmarksopen=false,breaklinks=false,pdfborder={0 0 0},backref=false,colorlinks=true,linkcolor=blue,citecolor=blue,urlcolor=blue]{hyperref}
\usepackage{tikz}

\newcommand{\dd}{\mathrm{d}}
\newcommand{\kb}[2]{\left| #1 \vphantom{#2} \right>\left< #2 \vphantom{#1} \right|} 
\newcommand{\proj}[1]{\kb{#1}{#1}} 
\newcommand{\eps}{\varepsilon}

\newtheorem{theo}{Theorem}

\begin{document}


\title{Constructing Local Models for General Measurements on Bosonic Gaussian States}


\author{Michael G. Jabbour}
\email{mgija@dtu.dk}
\affiliation{Department of Physics, Technical University of Denmark, 2800 Kongens Lyngby, Denmark}

\author{Jonatan Bohr Brask}
\email{jonatan.brask@fysik.dtu.dk}
\affiliation{Department of Physics, Technical University of Denmark, 2800 Kongens Lyngby, Denmark}


\begin{abstract}
We derive a simple sufficient criterion for the locality of correlations obtained from given measurements on a Gaussian quantum state. The criterion is based on the construction of a local-hidden-variable model that works by passing part of the inherent Gaussian noise of the state onto the measurements. We illustrate our result in the setting of displaced photodetection on a two-mode squeezed state. Here, our criterion exhibits the existence of a local-hidden-variable model for a range of parameters where the state is still entangled.
\end{abstract}

\maketitle

\textit{Introduction.}--Quantum mechanics allows for correlations than are impossible classically and which can be exploited in a variety of applications. In particular, entangled quantum states are a key resource for quantum information science, enabling advantages in computing, communication, and sensing \cite{Horodecki2009,Jozsa2003,Vidal2003,Giovanetti2006}. Furthermore, as shown by Bell \cite{Bell1964}, measurements on certain entangled states can lead to observations that violate a so-called Bell inequality and are then incompatible with local causal explanations. This phenomenon, known as nonlocality, demonstrates a profound departure from classical physics and is a cornerstone of modern understanding of quantum physics \cite{Brunner2014}. Nonlocal correlations also enable advantages for communication \cite{Cleve1997,Cubitt2011} and information processing at an unprecedented level of security \cite{colbeckPhD2009,Pironio09,Pironio2010}.

Entanglement and nonlocality, however, are not equivalent. While entanglement is a prerequisite for nonlocality, in general only carefully chosen measurements on a given entangled state will produce nonlocal observations, and while such measurements can always be found for pure entangled states \cite{Gisin1991}, there exist mixed entangled states that are local for any possible measurements \cite{Werner1989,Barrett2002}. Deciding whether given states can give rise to nonlocality is desirable both for applications and fundamentally.
This is, for instance, crucial in the context of device-independent (DI) quantum key distribution (QKD), the strongest form of quantum cryptographic protocols~\cite{Acin2007,Pironio2009}. In DIQKD and other DI protocols, security relies on the violation of a Bell inequality and hence requires the use of entangled states that enable nonlocality.

Certifying whether an entangled state exhibits nonlocality is far from trivial. To demonstrate nonlocality, it is sufficient to find a particular set of measurements that leads to violation of a particular Bell inequality. Demonstrating that a state cannot give rise to nonlocality is much harder because there are infinitely many possible measurements and Bell inequalities. It requires the construction of local-hidden-variable (LHV) models that can reproduce the observations for any combination of measurements. Constructing such models is challenging, even for particular classes of measurements. A number of methods for constructing LHV models have nevertheless been developed \cite{Werner1989,Popescu1994,Barrett2002,Almeida2007,Cavalcanti2016,Hirsch2016,Bowles2016,Fillettaz2018}, applicable to a variety of entangled states and measurements. Very often, a clear connection between the introduction of noise and the vanishing of nonlocality can be identified in these models, e.g., in~\cite{Werner1989,Almeida2007}.

While most previous work is concerned mainly with systems of finite dimension, another relevant class is that of so-called continuous-variables systems~\cite{Weedbrook2012}. Most particularly, Gaussian bosonic states and transformations are ubiquitous in quantum theory and in experiments in e.g.\ optical, superconducting, and mechanical platforms. At the same time, Gaussian systems are relatively easy to model. Their entanglement properties have been extensively studied \cite{Werner2001,Simon2000} and their nonlocality \cite{Banaszek1998,Banaszek1999,GarciaPatron2004,Revzen2005,Cavalcanti2007,Salles2008,Lee2009,He2010,Brask2012,Pozsgay2017} and steering \cite{Wiseman2007} have also been explored. The relation between noise and nonlocality has also been investigated \cite{Jeong2000,Mista2002}. For Gaussian measurements on Gaussian states, the resulting observations are always local, because the positive Wigner function of such states enables the construction of an LHV model for any set of Gaussian measurements (as explained in more detail below). However, little is known about the existence of LHV models for Gaussian states subject to non-Gaussian measurements.

Here, we develop a sufficient criterion for the existence of LHV models for general measurements on Gaussian states. Given a state and a candidate family of measurements, the criterion enables one to certify that they will never lead to nonlocal correlations.  The idea behind our result follows the lines of Werner and Wolf's criterion for the separability of Gaussian states~\cite{Werner2001}. Furthermore, we provide an interesting interpretation in terms of the role of noise for the vanishing of nonlocality, separating the inherent quantum noise resulting from the uncertainty relations from additional classical Gaussian noise. Before presenting our main result, we review some elements of the theory of bosonic systems and nonlocality.

\textit{Bosonic systems and Bell nonlocality.}--A bosonic system \cite{Weedbrook2012} is described by $N$ modes, where each mode is associated with an infinite-dimensional Hilbert space and a pair of bosonic field operators $\hat{a}_k, \hat{a}_k^{\dagger}$, where $k = 1,\ldots,N$ denotes the mode. The total system Hilbert space is the tensor product over the modes. The field operators satisfy the bosonic commutation relations $[\hat{a}_i,\hat{a}_j^\dag] = \delta_{ij}$, $[\hat{a}_i,\hat{a}_j] = 0$, $[\hat{a}_i^\dag,\hat{a}_j^\dag] = 0$. Alternatively, the system can be described using the quadrature operators $\{ \hat{q}_k, \hat{p}_k \}_{k=1}^N$ defined as $\hat{q}_k \coloneqq \hat{a}_k + \hat{a}_k^{\dagger}$, $\hat{p}_k \coloneqq i (\hat{a}_k^{\dagger} - \hat{a}_k)$ (we take $\hbar = 2$ throughout), which can also be arranged in the vector $\hat{\boldsymbol{r}} \coloneqq (\hat{q}_1, \hat{p}_1,\hdots,\hat{q}_N, \hat{p}_N)^{\mathrm{T}}$. The quadratures satisfy $[\hat{r}_k,\hat{r}_l] = 2 i \Omega_{kl}$, where $\boldsymbol{\Omega} \coloneqq \bigoplus_{k=1}^N \begin{pmatrix} 0 & 1 \\ -1 & 0 \end{pmatrix}$ is the symplectic form.

Any positive Hermitian operator in state space can equivalently be completely described by its real-valued Wigner function in phase space. If the operator is of unit trace (e.g.\ the density matrix $\rho$ of a quantum state), its Wigner function integrates to unity. Two quantities of particular interest are the two first statistical moments: the mean of the quadratures $\bar{\boldsymbol{r}} \coloneqq \mathrm{Tr}[\hat{\boldsymbol{r}} \hat{\rho}]$ and the covariance matrix $\boldsymbol{V}$ with $V_{ij} \coloneqq \mathrm{Tr}[\{ \Delta \hat{r}_i,\Delta \hat{r}_j \} \hat{\rho}]/2$, where $\Delta \hat{r}_i \coloneqq \hat{r}_i - \bar{r}_i$ and $\{ \cdot, \cdot \}$ is the anticommutator. Whenever $\rho$ is a genuine quantum state, the $2N \times 2N$ real, symmetric covariance matrix satisfies the uncertainty principle $\boldsymbol{V} + i \boldsymbol{\Omega} \geq 0$, which also implies $\boldsymbol{V} \geq 0$.

As already mentioned, Gaussian states \cite{Holevo1975} are ubiquitous in quantum experiments. These are states whose Wigner function is a multivariate Gaussian distribution. As such, they are completely described by their first two statistical moments, and their Wigner function can be written as
\begin{equation}
W(\boldsymbol{r}) = \frac{1}{(2 \pi)^N \sqrt{\det \boldsymbol{V}}} e^{-\frac{1}{2}(\boldsymbol{r}-\bar{\boldsymbol{r}})^\mathrm{T}\boldsymbol{V}^{-1}(\boldsymbol{r}-\bar{\boldsymbol{r}})}.    
\end{equation}
The entanglement in a Gaussian state is determined by its covariance matrix alone. A bipartite Gaussian state with covariance matrix $V_{AB}$ will be separable if and only if there exist genuine covariance matrices $\boldsymbol{\gamma}_A$ and $\boldsymbol{\gamma}_B$ of parties $A$ and $B$ such that $\boldsymbol{V} \geq \boldsymbol{\gamma}_A \oplus \boldsymbol{\gamma}_B$ \cite{Werner2001}.

A stronger form of correlations, Bell nonlocality is defined at the level of the observed input-output distribution in an experiment with multiple observers. In particular, a bipartite experiment with observers $A$ and $B$ is characterized by the distribution $p(ab|xy)$, where $x$, $y$ label the choice of input (measurement setting) of $A$ and $B$, respectively, and $a$, $b$ label their outputs (measurement outcomes).  The distribution is called nonlocal if it does not admit an LHV model, i.e., if it cannot be written as
\begin{equation} \label{eq:locmodel}
p(ab|xy) = \int \dd \lambda \, q(\lambda) p(a|x,\lambda) p(b|y,\lambda),
\end{equation} 
where the integral is over the (hidden) variable $\lambda$, which is distributed according to a probability density $q(\lambda)$ and where $p(a|x,\lambda)$ and $p(b|y,\lambda)$ are local response functions.

Entanglement is necessary but not sufficient for the generation of nonlocal correlations \cite{Brunner2014}. In a general bipartite quantum experiment, $A$ and $B$ share a state $\hat{\rho}_{AB}$ and each perform a generalized measurement with positive-operator-valued-measure (POVM) elements $Q_{a|x}$ and $R_{b|y}$, respectively. The corresponding probabilities are $p(ab|xy) = \mathrm{Tr}[\hat{\rho}_{AB} \, Q_{a|x} \otimes R_{b|y}]$. If the quantum state and all the POVM elements have positive Wigner functions, $p(ab|xy)$ is necessarily local. Indeed, if $\hat{\rho}_{AB}$, $Q_{a|x}$ and $R_{b|y}$ have respective Wigner functions $W$, $\mathcal{Q}_{a|x}$ and $\mathcal{R}_{b|y}$, we have
\begin{equation} \label{eq:lhvm}
    p(ab|xy) = \int \dd \boldsymbol{r} \, W(\boldsymbol{r}) \frac{\mathcal{Q}_{a|x}(\boldsymbol{r}_A)}{(4 \pi)^{-N_A}} \frac{\mathcal{R}_{b|y}(\boldsymbol{r}_B)}{(4 \pi)^{-N_B}} ,
\end{equation}
with $\boldsymbol{r} = (\boldsymbol{r}_A,\boldsymbol{r}_B)$, where $\boldsymbol{r}_A$ and $\boldsymbol{r}_B$ are the phase-space variables and $N_A$ and $N_B$ are the number of modes of party $A$ and $B$, respectively. This can be understood as an LHV model \eqref{eq:locmodel} with $\boldsymbol{r}$ as the hidden variable. $W$ is normalized and is hence a probability density over $\boldsymbol{r}$. Since $\sum_a Q_{a|x} = \mathbb{I}$, with $\mathbb{I}$ the identity operator, the Wigner functions fulfill $\sum_a \mathcal{Q}_{a|x}(\boldsymbol{r}_A) = (4 \pi)^{-N_A}$ for all $x$ and $\boldsymbol{r}_A$, because the Wigner function of the identity on $N$ modes is the constant $(4 \pi)^{-N}$ in our convention and similarly for $\mathcal{R}_{b|y}$. It follows that the last two terms in \eqref{eq:lhvm} are probability distributions over $a$ and $b$, respectively, and can be interpreted as local response functions. Hence \eqref{eq:lhvm} is of the form \eqref{eq:locmodel}. An immediate consequence is that correlations obtained by Gaussian measurements on a Gaussian state will never be nonlocal.

\textit{Constructing the LHV model}--We denote by $\mathcal{G}_{\bar{\boldsymbol{s}},\gamma}$ the multivariate Gaussian distribution with mean $\bar{\boldsymbol{s}}$ and covariance matrix $\gamma$, and by $f \ast g$ the convolution of functions $f$ and $g$, which is defined as
\begin{equation}
    (f \ast g)(\boldsymbol{r}) \coloneqq \int \dd \boldsymbol{r}' \, f(\boldsymbol{r}') g(\boldsymbol{r}-\boldsymbol{r}').
\end{equation}
We also define $\boldsymbol{0} \coloneqq (0,\hdots,0)^{\mathrm{T}}$. The following statement provides a sufficient criterion for the existence of LHV models for Gaussian states subject to specific measurements.
\begin{theo} \label{theo:main}
Let $\bar{\boldsymbol{r}}$ be the mean and $\boldsymbol{V}$ the covariance matrix of a Gaussian state $\hat{\rho}_{AB}$ and let $\mathcal{Q}_{a|x}$ and $\mathcal{R}_{b|y}$ be the Wigner functions of the POVM elements $Q_{a|x}$ and $R_{b|y}$. If there exist matrices $\gamma_A \geq 0$ and $\gamma_B \geq 0$ such that
\begin{equation} \label{eq:main1}
    \boldsymbol{V} \geq \gamma_A \oplus \gamma_B,
\end{equation}
and
\begin{equation} \label{eq:main2}
    \mathcal{Q}_{a|x} \ast \mathcal{G}_{\boldsymbol{0},\gamma_A} \geq 0 \quad \mathrm{and} \quad \mathcal{R}_{b|y} \ast \mathcal{G}_{\boldsymbol{0},\gamma_B} \geq 0,
\end{equation}
for all $a,x$ and $b,y$, then the probabilities $p(ab|xy) = \mathrm{Tr}[\hat{\rho}_{AB} \, Q_{a|x} \otimes R_{b|y}]$ exhibit an LHV model.
\end{theo}
\textit{Proof}. Let $\omega = \boldsymbol{V} - \gamma_A \oplus \gamma_B \geq 0$. Since $\gamma_A \geq 0$ and $\gamma_B \geq 0$, one can define genuine Gaussian probability distributions $\mathcal{G}_{\boldsymbol{0},\gamma_A}$ and $\mathcal{G}_{\boldsymbol{0},\gamma_B}$, and similarly for $\mathcal{G}_{\bar{\boldsymbol{r}},\omega}$. A useful property of Gaussian distributions is that convolving two such distributions results in a Gaussian distribution, i.e., $\mathcal{G}_{\bar{\boldsymbol{s}}_1,\gamma_1} \ast \mathcal{G}_{\bar{\boldsymbol{s}}_2,\gamma_2} = \mathcal{G}_{\bar{\boldsymbol{s}},\gamma}$, with $\bar{\boldsymbol{s}} = \bar{\boldsymbol{s}}_1 + \bar{\boldsymbol{s}}_2$ and $\gamma = \gamma_1 + \gamma_2$. Exploiting this and the symmetries of Gaussian distributions, we have
\begin{widetext}
\begin{equation} \label{eq:hvmproof}
    \begin{aligned}
    p(ab|xy) & = (4 \pi)^N \int \dd \boldsymbol{r}_A \dd \boldsymbol{r}_B \, \mathcal{G}_{\bar{\boldsymbol{r}},\boldsymbol{V}}(\boldsymbol{r}_A,\boldsymbol{r}_B) \mathcal{Q}_{a|x}(\boldsymbol{r}_A) \mathcal{R}_{b|y}(\boldsymbol{r}_B) \\
    & = (4 \pi)^N \int \dd \boldsymbol{r}_A \dd \boldsymbol{r}_B \, \int \dd \boldsymbol{r}'_A \dd \boldsymbol{r}'_B \, \mathcal{G}_{\bar{\boldsymbol{r}},\omega}(\boldsymbol{r}'_A,\boldsymbol{r}'_B)  \mathcal{G}_{\boldsymbol{0},\gamma_A \oplus \gamma_B}(\boldsymbol{r}_A-\boldsymbol{r}'_A,\boldsymbol{r}_B-\boldsymbol{r}'_B) \mathcal{Q}_{a|x}(\boldsymbol{r}_A) \mathcal{R}_{b|y}(\boldsymbol{r}_B) \\
    & = (4 \pi)^N \int \dd \boldsymbol{r}'_A \dd \boldsymbol{r}'_B \, \mathcal{G}_{\bar{\boldsymbol{r}},\omega}(\boldsymbol{r}'_A,\boldsymbol{r}'_B) \int \dd \boldsymbol{r}_A \dd \boldsymbol{r}_B \, \mathcal{G}_{\boldsymbol{0},\gamma_A \oplus \gamma_B}(\boldsymbol{r}'_A-\boldsymbol{r}_A,\boldsymbol{r}'_B-\boldsymbol{r}_B) \mathcal{Q}_{a|x}(\boldsymbol{r}_A) \mathcal{R}_{b|y}(\boldsymbol{r}_B) \\
    & = \int \dd \boldsymbol{r}_A \dd \boldsymbol{r}_B \, \mathcal{G}_{\bar{\boldsymbol{r}},\omega} (\boldsymbol{r}_A,\boldsymbol{r}_B) \frac{\tilde{\mathcal{Q}}_{a|x}(\boldsymbol{r}_A)}{(4 \pi)^{-N_A}} \frac{\tilde{\mathcal{R}}_{b|y}(\boldsymbol{r}_B)}{(4 \pi)^{-N_B}}
    \end{aligned}
\end{equation}
\end{widetext}
where $\tilde{\mathcal{Q}}_{a|x} \coloneqq \mathcal{Q}_{a|x} \ast \mathcal{G}_{\boldsymbol{0},\gamma_A} \geq 0$, $\tilde{\mathcal{R}}_{b|y} \coloneqq \mathcal{R}_{b|y} \ast \mathcal{G}_{\boldsymbol{0},\gamma_B} \geq 0$. Since for the constant distribution $c=(4 \pi)^{-N_A}$, it holds that $c \ast \mathcal{G}_{\boldsymbol{0},\gamma_A} = c$,
we also have that $(4 \pi)^{N_A} \sum_{a} \tilde{\mathcal{Q}}_{a|x} = 1$, and similarly for $\tilde{\mathcal{R}}_{b|y}$. Eq.~\eqref{eq:hvmproof} can therefore be interpreted as an LHV model.~$\square$

An important point of Theorem~\ref{theo:main} is that $\gamma_A$ and $\gamma_B$ need not be covariance matrices of genuine quantum states, and only have to be non-negative. It is instructive to have a closer look at the situation when the state $\hat{\rho}_{AB}$ is separable. In that case, there exist covariance matrices $\gamma_A$ and $\gamma_B$ of quantum states (i.e., which satisfy the uncertainty principle), such that $\boldsymbol{V} \geq \gamma_A \oplus \gamma_B$~\cite{Werner2001}, so that
\begin{equation}
    \begin{aligned}
        \tilde{\mathcal{Q}}_{a|x}(\boldsymbol{r}_A) & = \int \dd \boldsymbol{s}_A \, \mathcal{Q}_{a|x}(\boldsymbol{s}_A) \mathcal{G}_{\boldsymbol{0},\gamma_A}(\boldsymbol{r}_A-\boldsymbol{s}_A) \\
        & = \int \dd \boldsymbol{s}_A \, \mathcal{Q}_{a|x}(\boldsymbol{s}_A) \mathcal{G}_{\boldsymbol{r}_A,\gamma_A}(\boldsymbol{s}_A) \\
        & = (4 \pi)^{-N_A} \, \mathrm{Tr} \left[ Q_{a|x} \hat{\sigma}_A \right],
    \end{aligned}
\end{equation}
where $\hat{\sigma}_A$ is the density matrix of the Gaussian state with mean value $\boldsymbol{r}_A$ and covariance matrix $\gamma_A$. Since $\hat{\sigma}_A$ is a genuine density matrix, we have that $\tilde{\mathcal{Q}}_{a|x}(\boldsymbol{r}_A) \geq 0$ for all $\boldsymbol{r}_A$. The same reasoning can, of course, be made for party $B$. We therefore see that, when $\hat{\rho}_{AB}$ is separable, we are always provided with an LHV model whatever the measurements, as should indeed be the case.

In fact, while the Wigner functions $\tilde{\mathcal{Q}}_{a|x}$ and $\tilde{\mathcal{R}}_{b|y}$ will always become positive when subject to enough noise (that is, noise coming from a separable state $\hat{\rho}_{AB}$), one can push the analysis further. Consider the bivariate convolution $\tilde{\mathcal{Q}}^{(t)}_{a|x} \coloneqq \mathcal{Q}_{a|x} \ast \mathcal{G}_{\boldsymbol{0},\gamma_A}$ with the choice $\gamma_A = t \mathbb{I}_2$, for some $t \geq 0$, where $\mathbb{I}_2$ is the $2\times 2$ identity matrix. It is well known that the function $\tilde{\mathcal{Q}}^{(t)}_{a|x}$ then satisfies the heat (or diffusion) equation~\cite{Mathews1970}
\begin{equation} \label{eq:heat}
    \frac{\partial}{\partial t} \tilde{\mathcal{Q}}^{(t)}_{a|x} = \frac{1}{2} \Delta \tilde{\mathcal{Q}}^{(t)}_{a|x},
\end{equation}
where $\Delta$ is the Laplacian, with initial condition $\tilde{\mathcal{Q}}^{(0)}_{a|x} = \mathcal{Q}_{a|x}$.
In the limit of $t \rightarrow \infty$, the function $\tilde{\mathcal{Q}}^{(t)}_{a|x}$ approaches a Gaussian, which is necessarily non-negative everywhere. The convolution $\mathcal{Q}_{a|x} \ast \mathcal{G}_{\boldsymbol{0},\gamma_A}$ actually always makes the quasiprobability distribution $\mathcal{Q}_{a|x}$ ``less negative" as the parameter $t$ increases. More precisely, the local minima of $\tilde{\mathcal{Q}}^{(t)}_{a|x}$ have non-negative Laplacian, which implies from the heat equation~\eqref{eq:heat} that their $t$ derivative is non-negative, so that their values never decrease when $t$ increases.

One can give an operational interpretation of Theorem~\ref{theo:main} in terms of the effect that added local Gaussian noise has on nonlocality. Consider a bipartite pure Gaussian state $\hat{\rho}_{AB}$ with covariance matrix $\boldsymbol{V}$ and suppose it can be written as $\boldsymbol{V} = \omega + \gamma^{\mathrm{q}}_A \oplus \gamma^{\mathrm{q}}_B$ with $\omega, \gamma^{\mathrm{q}}_A, \gamma^{\mathrm{q}}_B \geq 0$ (where q is for quantum). Suppose further that we apply local noise to $\hat{\rho}_{AB}$ in the form of classical additive Gaussian noise channels~\cite{Weedbrook2012}, i.e., local quantum convolutions in the sense of Ref.~\cite{Werner1984}. These channels are completely characterized by their action on the covariance matrix, which is of the form $\boldsymbol{V} \mapsto \boldsymbol{V} + \gamma^{\mathrm{c}}_A \oplus \gamma^{\mathrm{c}}_B$ with $\gamma^{\mathrm{c}}_A, \gamma^{\mathrm{c}}_B \geq 0$ (where c is for classical). The resulting mixed Gaussian state $\hat{\rho}_{AB}'$ has covariance matrix $\omega + (\gamma^{\mathrm{q}}_A + \gamma^{\mathrm{c}}_A) \oplus (\gamma^{\mathrm{q}}_B + \gamma^{\mathrm{c}}_B)$. Now apply Theorem~\ref{theo:main} to $\hat{\rho}_{AB}'$ with the POVM elements $Q_{a|x}$ and $R_{b|y}$. An LHV model will exist if 
\begin{equation} \label{eq:covqc}
    \mathcal{Q}_{a|x} \ast \mathcal{G}_{\boldsymbol{0},\gamma^{\mathrm{q}}_A + \gamma^{\mathrm{c}}_A} \geq 0 \quad \mathrm{and} \quad \mathcal{R}_{b|y} \ast \mathcal{G}_{\boldsymbol{0},\gamma^{\mathrm{q}}_B + \gamma^{\mathrm{c}}_B} \geq 0,
\end{equation}
for all $a,x$ and $b,y$. Eq~\eqref{eq:covqc} expresses the fact that $\hat{\rho}_{AB}'$ will become local with respect to the POVMs $Q_{a|x}$ and $R_{b|y}$ when the noise provided by the convolutions with the Gaussian distributions $\mathcal{G}_{\boldsymbol{0},\gamma^{\mathrm{q}}_A + \gamma^{\mathrm{c}}_A}$ and $\mathcal{G}_{\boldsymbol{0},\gamma^{\mathrm{q}}_B + \gamma^{\mathrm{c}}_B}$ is important enough. There are two contributions to the noise. The first, characterized by $\gamma^{\mathrm{q}}_A$ and $\gamma^{\mathrm{q}}_B$, is quantum noise; that is, the uncertainty inherent to quantum mechanics coming from the fact that the pure state $\hat{\rho}_{AB}$ is subject to the uncertainty relation. The second, characterized by $\gamma^{\mathrm{c}}_A$ and $\gamma^{\mathrm{c}}_B$, is classical Gaussian additive noise making the state mixed. An interesting situation arises when either $\gamma^{\mathrm{q}}_A + \gamma^{\mathrm{c}}_A$ or $\gamma^{\mathrm{q}}_B + \gamma^{\mathrm{c}}_B$ is not a genuine covariance matrix, so that $\hat{\rho}_{AB}'$ is still entangled, while the noise is important enough so that there exists an LHV model. We provide an example of this in the following.

\begin{figure}[t]
\centering
	\includegraphics[width=.8\columnwidth]{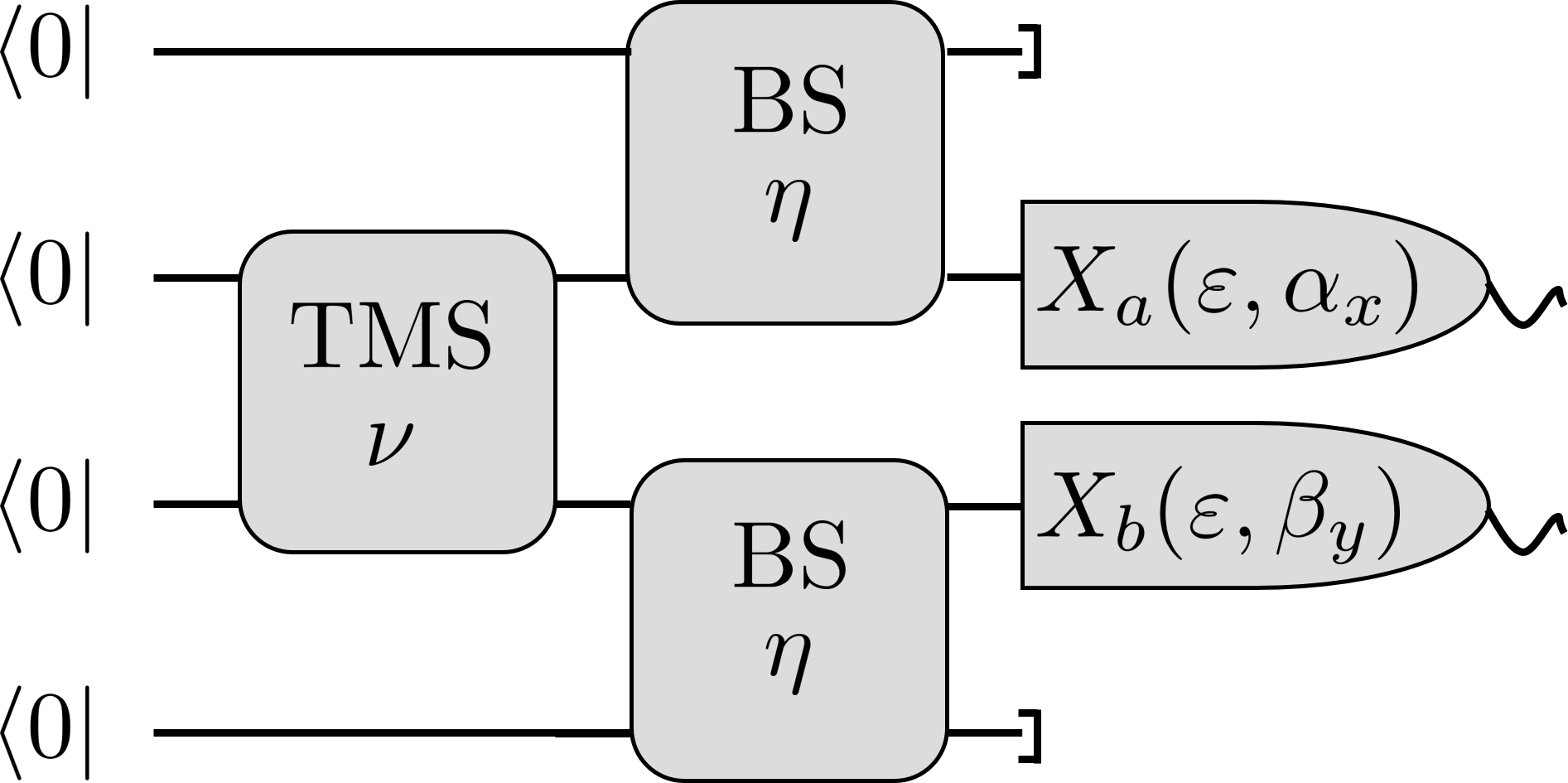}
	\caption{\label{fig:Bell} Sketch of a scheme for demonstrating nonlocality from a TMSS with loss. A TMSS is generated by injecting a couple of vacua into a two-mode squeezer (TMS) of parameter $\nu$, while losses are modeled by the interaction of each output mode of the TMS with a vacuum state through a beam splitter (BS) of transmittance $\eta$. The measurements characterized by the POVM elements $X_a(\eps,\alpha_x)$ and $X_b(\eps,\beta_y)$ are then performed on party $A$ and $B$, respectively.}
\end{figure}

\textit{An application.}--For the sake of illustration, we consider a two-mode squeezed state (TMSS) $\hat{\rho}_{AB}$ with zero mean and covariance matrix
\begin{equation} \label{eq:covTMSS}
    \boldsymbol{V} = \begin{pmatrix} \nu \mathbb{I}_2 & \sqrt{\nu^2-1} \boldsymbol{Z} \\ \sqrt{\nu^2-1} \boldsymbol{Z} & \nu \mathbb{I}_2 \end{pmatrix},
\end{equation}
where $\nu \geq 1$ and $\boldsymbol{Z} \coloneqq \mathrm{diag}(1,-1)$. It is entangled for $\nu > 1$. We consider a scheme similar to that of Ref.~\cite{Brask2012} for demonstrating nonlocality with a TMSS (see Fig.~\ref{fig:Bell}). First, we take losses into account by applying a local pure-loss channel~\cite{Weedbrook2012} $\mathcal{E}_\eta$ of parameter $\eta \in [0,1]$ to each mode of the TMSS. The channel $\mathcal{E}_\eta$ acts as
\begin{equation}
    \mathcal{E}_\eta[\sigma] := \mathrm{Tr}_2 \left[ U_\eta \left( \sigma \otimes \proj{0} \right) U_\eta^\dagger \right],
\end{equation}
where $U_\eta$ is a beam-splitter unitary and $\ket{0}$ is the vacuum state. Since $\mathcal{E}_\eta$ is Gaussian, the resulting state $\hat{\rho}'_{AB} = \left( \mathcal{E}_\eta \otimes \mathcal{E}_\eta \right)[\hat{\rho}_{AB}]$ is also Gaussian with zero mean value and covariance matrix
\begin{equation} \label{eq:covTMSSeta}
    \boldsymbol{V}' = \begin{pmatrix} [1+\eta(\nu-1)] \mathbb{I}_2 & \eta \sqrt{\nu^2-1} \boldsymbol{Z} \\ \eta \sqrt{\nu^2-1} \boldsymbol{Z} & [1+\eta(\nu-1)] \mathbb{I}_2 \end{pmatrix}.
\end{equation}
Furthermore, it can be seen to be entangled for any $\nu>1$ and $\eta>0$ by evaluating the partial transpose \cite{Peres1996,Horodecki1996,Simon2000}.

Next, for the measurements we consider displacements followed by non-number-resolving single-photon detection (click or no-click). Ideally, this implements a measurement where the no-click outcome corresponds to a projection onto a coherent state. Here, we allow for some noise in the detection by modeling the POVM element corresponding to the no-click outcome as $X_{+1}(\eps,\alpha) \coloneqq D_{\alpha} [(1-\eps)\proj{0} + \eps \proj{1}] D_{\alpha}^{\dagger}$, where $D_{\alpha}$ is the displacement operator and $\ket{1}$ is the one-photon Fock state. The click outcome corresponds to $X_{-1}(\eps,\alpha) \coloneqq \mathbb{I} - X_{+1}(\eps,\alpha)$. The parameter $\epsilon\in[0,1]$ can be understood as the probability for an additional excitation to be introduced during measurement. 

\begin{figure}[t]
\centering
	\includegraphics[width=.95\columnwidth,trim=4 4 4 4]{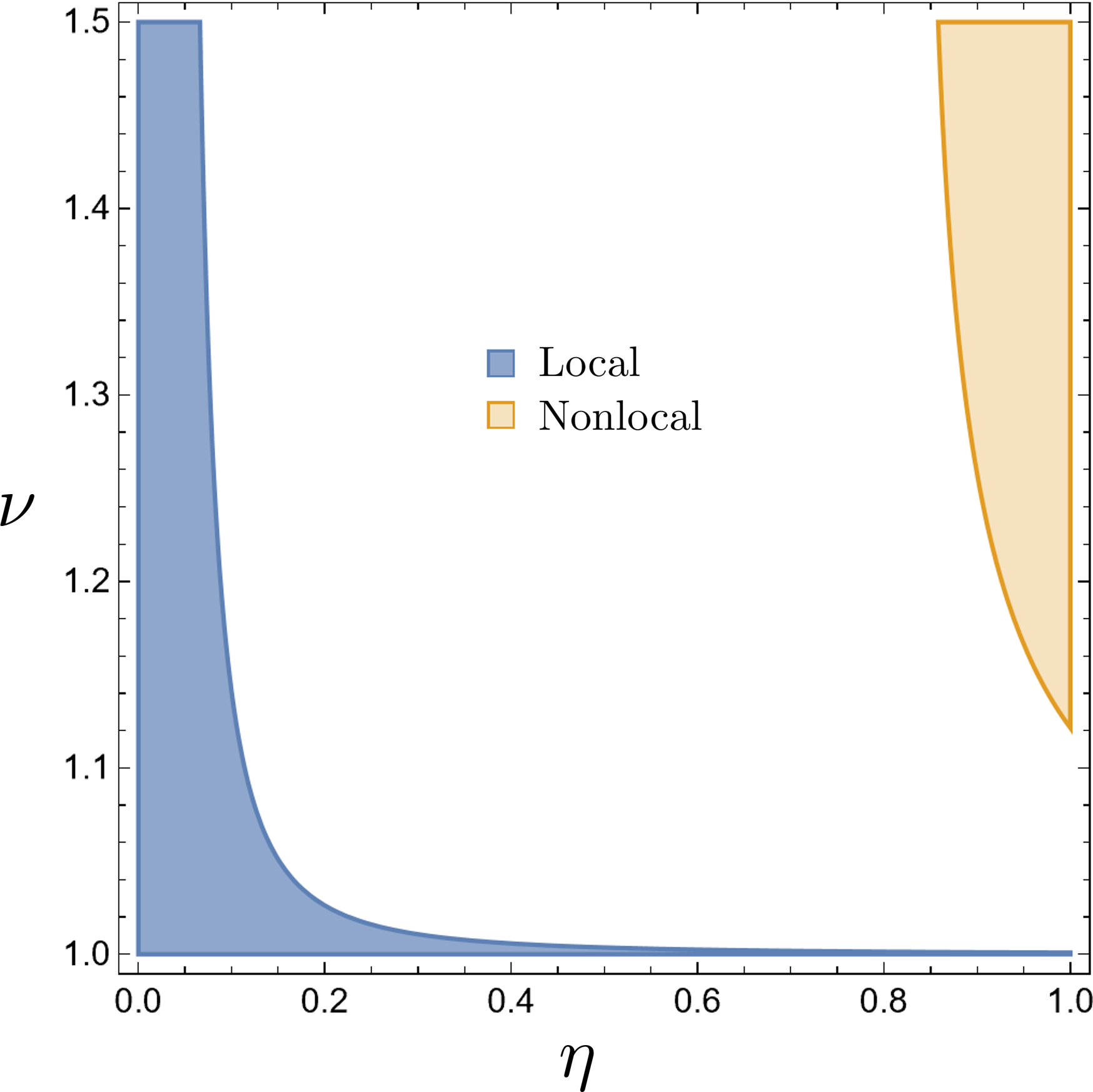}
	\caption{\label{fig:chsh} Comparisons of regions for which the lossy TMSS $\hat{\rho}'_{AB}$ admits an LHV model for the choice of measurements $\{X_{+1}(\eps,\alpha),X_{-1}(\eps,\alpha)\}$ (blue region, left) and for which it violates the CHSH inequality (orange region, right), for the choice $\eps=0.02$. Note that we limited the figures to values $\nu \in [1,1.5]$ as this is where a significant violation of the CHSH inequality occurs, but the region of existence of LHV models extends further when increasing the range of values of $\nu$.}
\end{figure}

Inputs $x,y\in\{0,1\}$ for $A$ and $B$ correspond to displacements $\alpha_x$ and $\beta_y$, respectively, and we label the outputs $a,b\in\{\pm 1\}$, with $-1$ for click events. We take the noise strength $\eps$ to be the same for all measurements. Defining the correlators $\left\langle a_{x} b_{y} \right\rangle = \sum_{a,b} ab \, p(ab|xy)$, Eq.~\eqref{eq:locmodel} implies the Clauser-Horne-Shimony-Holt (CHSH) inequality~\cite{Clauser1969}
\begin{equation} \label{eq:chsh}
    S = \left\langle a_{0} b_{0} \right\rangle + \left\langle a_{0} b_{1} \right\rangle + \left\langle a_{1} b_{0} \right\rangle - \left\langle a_{1} b_{1} \right\rangle \leq 2 .
\end{equation}
This inequality can be violated for the quantum probabilities $p(ab|xy) = \mathrm{Tr}[\hat{\rho}'_{AB} (X_a(\eps,\alpha_x) \otimes X_b(\eps,\beta_y))]$. In particular, taking $\beta_y = - \alpha_x$ for $y=x$ and optimizing over real $\alpha_x$, we find violation for a range of values of the squeezing, loss, and noise, as shown in Fig.~\ref{fig:chsh}. To do so, we fix $\eps=0.02$ as an example, before numerically maximizing the value of $S$ in Eq.~\eqref{eq:chsh} over the free parameters $\alpha_0, \alpha_1 \in [-1,1]$, for each value of $\eta \in [0,1]$ and $\nu \in [1,1.5]$, after a suitable discretization. For instance, if one chooses $\eta = 0.95$ and $\nu = 1.4$, one gets $S \simeq 2.1 > 2$ for $(\alpha_0, \alpha_1) \simeq (0.12,-0.48)$.

On the other hand, we can apply Theorem~\ref{theo:main} to show that the correlations must be local for another parameter region. Let $\mathcal{X}_a^{(\eps,\alpha)}$ be the Wigner function of $X_a(\eps,\alpha)$. The quasidistribution $\mathcal{X}_{+1}^{(\eps,\alpha)}$ is non-negative everywhere since $\eps<1$, while $\mathcal{X}_{-1}^{(\eps,\alpha)}$ admits negative values. According to Theorem~\ref{theo:main}, the probability $p(ab|xy)$ will satisfy Eq.~\eqref{eq:locmodel} if there exist non-negative matrices $\gamma_A$ and $\gamma_B$ such that $\boldsymbol{V}' \geq \gamma_A \oplus \gamma_B$ and the Wigner functions $\mathcal{X}_a^{(\eps,\alpha_x)} \ast \mathcal{G}_{\boldsymbol{0},\gamma_A}$ and $\mathcal{X}_b^{(\eps,\beta_y)} \ast \mathcal{G}_{\boldsymbol{0},\gamma_B}$ are non-negative for all $a,b$. It is enough to find $\gamma_A$ and $\gamma_B$ such that $\mathcal{X}_{-1}^{(\eps,\alpha_x)} \ast \mathcal{G}_{\boldsymbol{0},\gamma_A} \geq 0$ and $\mathcal{X}_{-1}^{(\eps,\beta_y)} \ast \mathcal{G}_{\boldsymbol{0},\gamma_B} \geq 0$. Now, consider the choice $\gamma_A = \gamma_B = t \mathbb{I}_2$ with $t \geq 0$. If we are to satisfy $\boldsymbol{V}' \geq \gamma_A \oplus \gamma_B$, we need $t \leq 1 + \eta (\nu -1 - \sqrt{\nu^2-1})$. From Eq.~\eqref{eq:heat}, it follows that if $\mathcal{X}_{-1}^{(\eps,\alpha)} \ast \mathcal{G}_{\boldsymbol{0},t \mathbb{I}_2}$ becomes non-negative for some value of $t$, it will remain so for all larger $t$. Consequently, one can consider the highest acceptable value of $t$, that is $t = 1 + \eta (\nu -1 - \sqrt{\nu^2-1})$. Furthermore, by definition of the convolution, the value of $t$ for which $\mathcal{X}_{-1}^{(\eps,\alpha)} \ast \mathcal{G}_{\boldsymbol{0},t \mathbb{I}_2}$ becomes non-negative does not depend on $\alpha$, so that one can take $\alpha=0$.
Now, the Wigner functions of the operators $\mathbb{I}$, $\proj{0}$ and $\proj{1}$ are, respectively, given by
$W_{\mathbb{I}}(x,p) = 1/(4 \pi)$,
$W_{\proj{0}}(x,p) = e^{-(x^2+p^2)/2}/(2\pi)$ and
$W_{\proj{1}}(x,p) = -(1-x^2-p^2) e^{-(x^2+p^2)/2}/(2\pi)$~\cite{leonhardt2010}, while we have $\mathcal{X}_{-1}^{(\eps,\alpha)}(x,p) = W_{\mathbb{I}} - (1-\eps) W_{\proj{0}}(x,p) - \eps W_{\proj{1}}(x,p)$. Using this, we obtain
\begin{equation}
    \begin{aligned}
    & (\mathcal{X}_{-1}^{(\eps,0)} \ast \mathcal{G}_{\boldsymbol{0},t \mathbb{I}_2})(x,p) \\
    & = \frac{1}{4 \pi} - \frac{(1+t)^2+\eps[x^2+p^2-2(1+t)]}{2 \pi (1+t)^3} e^{\frac{-(x^2+p^2)}{2(1+t)}}.
    \end{aligned}
\end{equation}
It can easily be shown that for all $t \geq 0$ and $(x,p) \in \mathbb{R}^2$, the above function achieves its minimum at $(x,p) = 0$, and that this minimum is non-negative for $t \geq \sqrt{1-4\eps}$. From our choice of $t$, this means that the distribution will be non-negative everywhere for $1 + \eta (\nu -1 - \sqrt{\nu^2-1}) \geq \sqrt{1-4\eps}$. The corresponding region in the $(\eta,\nu)$-plane is plotted in Fig.~\ref{fig:chsh} for $\eps=0.02$. In this region, the entangled state $\hat{\rho}'_{AB}$ admits an LHV model for the family of measurements described above.

\textit{Conclusion.}--In this work, we have developed a criterion for the existence of local-hidden-variable models for correlations resulting from general measurements on Gaussian states, by exploiting that measurement-operator Wigner functions can be made positive by passing Gaussian noise from the state to the measurement. We have illustrated the criterion for the case of noisy displacement-based measurements on a two-mode squeezed state subject to loss.

Recently, continuous-variables quantum systems have emerged as a promising platform for the implementation of QKD protocols~\cite{Ralph2000,Grosshans2002,Weedbrook2004,Jouguet2013,Ye2020,Ghalaii2022}. In particular, Gaussian systems such as coherent states can serve as a resource for security against collective attacks~\cite{Grosshans2005,Renner2009}. In light of this, we expect the present work to also be useful in the context of DIQKD with Gaussian states.

An interesting question is whether the statement of Theorem~\ref{theo:main} is also a necessary criterion: if one cannot find two positive semidefinite matrices $\gamma_A$ and $\gamma_B$ such that Eq.~\eqref{eq:main2} is satisfied for all POVM elements simultaneously, does this imply nonlocality of the distribution $p(ab|xy) = \mathrm{Tr}[\hat{\rho}_{AB} Q_{a|x} \otimes R_{b|y}]$?

\begin{acknowledgments}
We gratefully acknowledge support from the Carlsberg Foundation CF19-0313, the Independent Research Fund Denmark 7027-00044B, and VILLUM FONDEN grant 40864.
\end{acknowledgments}


\bibliography{nonlocalityGaussian}

\end{document}